\documentclass[amsmath,amssymb,aps,pre,nofootinbib,showkeys]{revtex4-2}
\usepackage{amsmath}
\usepackage{graphicx}
\usepackage{booktabs, threeparttable, cellspace, tabularx}
\usepackage{natbib}
\usepackage{bbm}
\usepackage{hhline}
\usepackage{color, colortbl}
\graphicspath{ {Images/} }
\usepackage{caption}
\usepackage{array}
\newcolumntype{C}{>{\centering\arraybackslash}X}
\newcolumntype{L}{>{\raggedright\arraybackslash}X}
\newcolumntype{R}{>{\raggedleft\arraybackslash}X}
\newcolumntype{P}[1]{>{\centering\arraybackslash}p{#1}}
\begin{document}
	
	\title{A complex networks approach to ranking professional Snooker players}
	\author{Joseph D.~\surname{O'Brien}}
	\email{joseph.obrien@ul.ie}
	\affiliation{MACSI, Department of Mathematics and Statistics, University of Limerick, Limerick V94 T9PX, Ireland}
	\author{James P.~Gleeson}
	\affiliation{MACSI, Department of Mathematics and Statistics, University of Limerick, Limerick V94 T9PX, Ireland}
	
	\begin{abstract}
		A detailed analysis of matches played in the sport of Snooker during the period 1968-2020 is used to calculate a directed and weighted dominance network based upon the corresponding results. We consider a ranking procedure based upon the well-studied PageRank algorithm that incorporates details of not only the number of wins a player has had over their career but also the quality of opponent faced in these wins. Through this study we find that John Higgins is the highest performing Snooker player of all time with Ronnie O'Sullivan appearing in second place. We demonstrate how this approach can be applied across a variety of temporal periods in each of which we may identify the strongest player in the corresponding era. This procedure is then compared with more classical ranking schemes. Furthermore, a visualization tool known as the rank-clock is introduced to the sport which allows for immediate analysis of the career trajectory of individual competitors. These results further demonstrate the use of network science in  the quantification of success within the field of sport.
	\end{abstract}
	\maketitle
	\section{Introduction}
	
	Each day competitive contests between similar entities occur in the hope of one proving dominant over the other. These contests have been shown to be wide-ranging with examples including animals combating in order to prove their strength~\cite{Chase2002, Ellis2017}, online content producers aiming to create a popular post~\cite{Weng2012, Gleeson2014, Lorenz-Spreen2019}, or the quantification of the scientific quality underlying a researcher's output~\cite{Lehmann2006, Radicchi2008, Radicchi2009, Sinatra2016}. In most of these scenarios it proves difficult to ultimately determine the stronger of two such competitors for a number of reasons, most evidently  the lack of explicit quantitative data describing the corresponding result from each contest. One noticeable exemption to this predicament is in the case of competitive sports where, on the contrary, there exists an abundance of data available from extended periods of time describing the results of  contests. This source of empirical data has resulted in an entire domain of study in applying the theoretical concepts of complex systems to the field of sport~\cite{Passos2011, Davids2013, Wasche2017}. 
	
	The application of these tools has resulted in a greater understanding of the dynamics underlying a number of sporting contests including soccer~\cite{Grund2012, Buldu2019}, baseball~\cite{Petersen2008, Saavedra2010},  basketball~\cite{Gabel2012, Clauset2015, Ribeiro2016}, and more recently even virtual sporting contests based upon actual sports~\cite{Getty2018, OBrien2020}. An area that has received much focus and which is most relevant to the present work is the application of network science~\cite{Newman2010} in identifying important sporting competitors in both team and individual sports. This has led to analysis in a range of sports including team-based games such as soccer where the identification of important players within a team's structure has been considered~\cite{Onody2004, Duch2010} and cricket, where rankings of both teams and the most influential player in specialty roles including captains, bowlers, and batsmen have been considered~\cite{Mukherjee2012, Mukherjee2014}. Analysis has also been conducted into individual-based sports, again with the aim of providing a ranking of players within a given sport. For example, the competitors within both professional tennis~\cite{Radicchi2011} and boxers, at an individual weight level~\cite{TennantBoxing17} and a pound-for-pound level~\cite{Tennant20}, have been extensively studied. Lastly, rankings at a country level based upon their success across the spectrum of Olympic Games sports  have also been considered~\cite{Calzada-Infante2016}.
	
	In this article we focus on the application of network science to the sport of Snooker --- a cue-based game with its origins in the late $19^{\text{th}}$ century from the military bases of British officers based in India. The game is played on a cloth-covered rectangular table which has six pockets located at the four corners and two along the middle of the longer sides. The players strike the cue ball (which is white) with their cue such that it strikes another of the 21 colored balls which is then ideally pocketed ,i.e., it falls into one of the pockets. The order in which the different colored balls must be pocketed is pre-determined and a player is awarded different points depending on the color of ball pocketed. For each set of balls (known as a \textit{frame}) one player has the first shot and continues to play until they fail to pocket a ball, at which point their competitor then has their own attempt. The number of points scored by a player in a single visit to the table is known as a \textit{break}. The player who has the most points after all balls have been pocketed (or the other player concedes) is the winner of the frame. A snooker match itself general consists of an odd number of frames such that the players compete until it is impossible for the other to win , i.e., the winner reaches a majority of frames.
	
	The popularization of Snooker came in conjunction with the advent of color television where the sport  demonstrated the potential applicability of this new technology for entertainment purposes~\cite{Bury1986}. From the 1970s onwards Snooker's popularity grew among residents of the United Kingdom and Ireland, culminating in the 1985 World Championship~---~the final of which obtained a viewership of 18.5 million, a record at the time for any broadcast shown after midnight in the United Kingdom. The sport continued to increase in popularity over the 1990s with there being a significant increase in the number of professional players. It was, however, dealt a blow in 2005 with the banning of sponsorship from the tobacco companies who were major benefactors of the sport. There has been a revitalization 
	in the sport over the past decade however with a reorganization of the governing body \textit{World Snooker}~\cite{worldsnooker} and a resulting increase in both the number of competitive tournaments alongside the corresponding prize-money on offer for the players. One notable consequence of this change is the way in which the official rankings of the players is determined. Specifically, there has been a change from the points based system which was the method of choice from 1968 to 2013 towards a system based upon the player's total prize-winnings in monetary terms. The question arises as to whether this approach more accurately captures the actual ranking of a player's performances over the season in terms of who is capabable of beating whom and if, alternatively, a more accurate approach exists. 
	
	Motivated by this question, in this paper we consider a dataset of competitive Snooker matches taken over a period of over fifty years (1968 - 2020) with the aim of firstly constructing a networked representation of the contests between each player. This is obtained by representing all the matches between two players as a weighted connection, which we show to have similar features to apparently unrelated complex systems~\cite{Newman2010}. Using this conceptual network we proceed to make use of a ranking algorithm similar in spirit to the PageRank algorithm~\cite{Gleich2015} from which we can quantify the quality of players over multiple different temporal periods within our dataset. Importantly this algorithm is based purely on the network topology itself and does not incorporate any exogenous factors such as points or prize-winnings. The benefit offered by this approach is that, through the aforementioned dominance network, the quality of competition faced in each game is incorporated when determining a player's rank rather than simply the final result itself. As such, the algorithm places higher levels of importance in victory against other players who are perceived as successful and, notably, similar approaches have proven effective when applied to other sports resulting in numerous new insights to the competitive structure of said contests~\cite{Radicchi2011, Mukherjee2012, Mukherjee2014, TennantBoxing17, Tennant20}. Through this ranking system we proceed to highlight a number of interesting properties underlying the sport including an increased level of competition among players over the previous thirty years. We also demonstrate that while prior to its revitalization Snooker was failing to capture the dominance-ranking of players in the sport through its points-based ranking scheme, the subsequent change in the ranking system to a prize-money basis is also inaccurate 
	We investigate the quality of different ranking schemes in comparison to the PageRank approach using similarity metrics and also introduce a graphical tool known as a rank-clock~\cite{Batty2006} to the sport of Snooker which allows one to interpret how a player's rank has changed over the course of their career. Finally, we conclude with a discussion on the work and how it offers the potential for a new form of ranking scheme within the sport of Snooker.
	
	\section{Methods}
	
	The data used in the analysis to follow is obtained from the \textit{cuetracker} website \cite{cuetracker} which is an online database containing information of professional snooker tournaments from 1908 onwards. This amounts to providing the records of 18,324 players from a range of skill levels. In this article we focus only on those matches which were of a competitive professional nature and took place between the years of 1968 and 2020, a period of over fifty years. More specifically, in terms of the quality of match considered, we focus on those games that fall under the categories League, Invitational and Ranking events which are those that the majority of professional players compete in. With these considerations the dataset used amounts to 657 tournaments featuring 1221 unique players competing in 47710 matches. Importantly each season is split over two years such that the season which begins in one calendar year concludes during the following calendar year. As such we reference seasons by the year in which they begun i.e., the 2018-19 season is referred to as the 2018 season in analysis below. Furthermore, for validation and comparative purposes in the forthcoming analysis, the official rankings of snooker players from World Snooker (the governing body of the sport) were also obtained~\cite{worldsnooker} from the period 1975-2020. The top two panels of Fig.~\ref{fig:summaries} demonstrate the temporal behavior behind both the number of players and tournaments in each season of our dataset. We see the increase in popularity of Snooker and corresponding financial sustainability for more players arising during the period 1980-2000 prior to the subsequent decrease in professionals and tournament in the decade to follow. In the concluding ten years of the dataset however we observe an increase in both the number of tournaments alongside the players who compete in them as a possible consequence of the professional game's restructuring.
	
	\subsection{Network Generation}\label{subsec:network}
\begin{figure*}
		\centering
		\includegraphics[width=\textwidth]{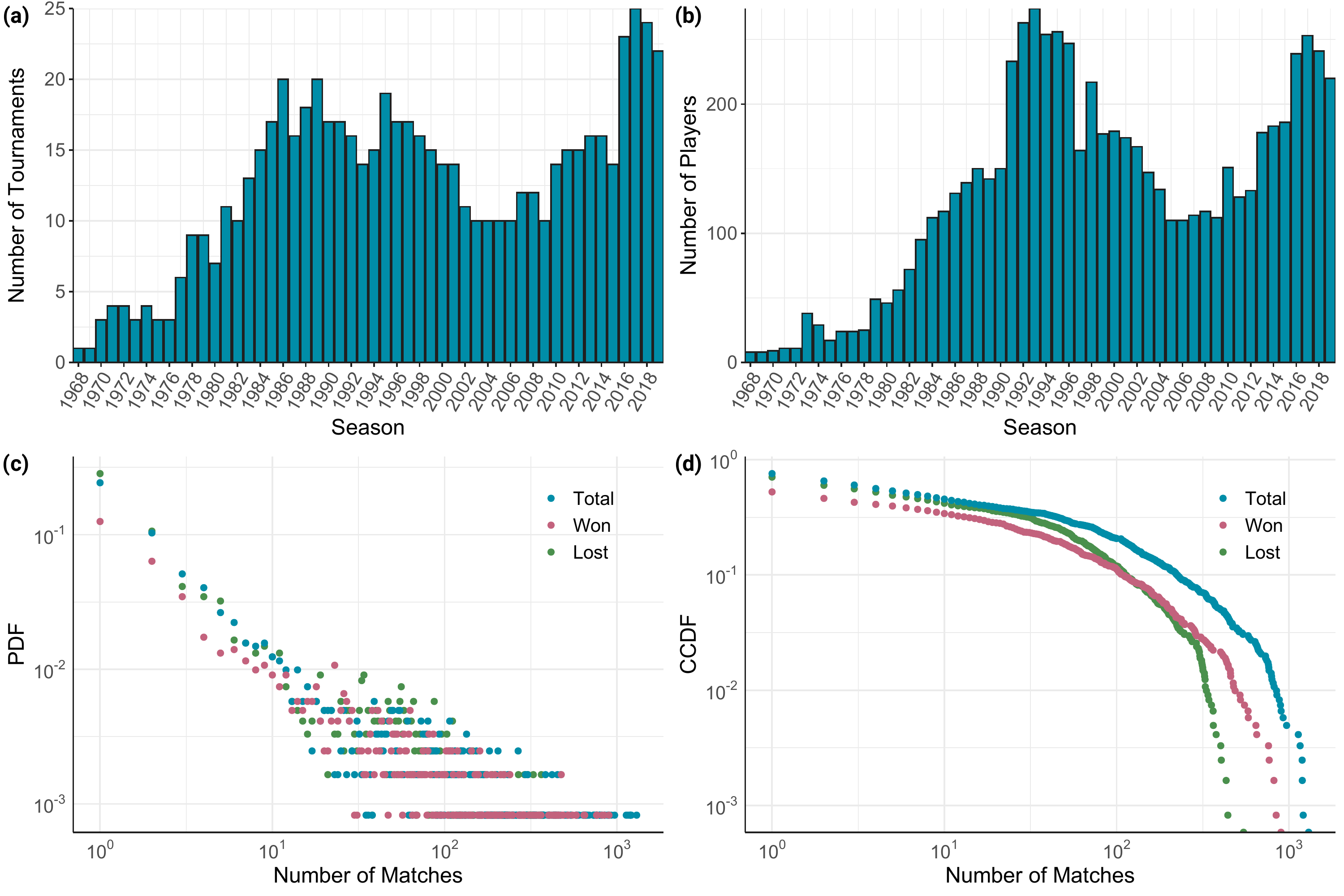}
		\caption{\textbf{Summary of the Snooker result dataset.} \textbf{(a)}~The total number of tournaments taking place in each season, the increasing popularity of the sport over the period 1980-2000 can be seen with increased number of tournaments and similarly its subsequent decay until the second half of the last decade. \textbf{(b)} Similar to panel (a) but now showing the number of professional players who competed by year in our dataset. \textbf{(c)}~The probability distribution function~(PDF) of player results describing the fraction of players who have played~(blue), won~(red), and lost~(green) a certain number of games. Each of these quantities appear to follow a heavy-tailed distribution. \textbf{(d)}~The corresponding complementary cumulative distribution function~(CCDF) in each case of panel (c), i.e., the fraction of players that have more than a certain number of matches in each category.}
		\label{fig:summaries}
	\end{figure*}

	In order to create a networked representation of the competition between pairs of Snooker players we consider each match in our dataset as an edge of weight one and construct a dominance relationship between the two players appearing in said match. Thus each time that player~$i$ defeats player~$j$, an edge from~$j$ to~$i$ is drawn. The construction described above results in a directed, weighted network with entries~$w_{ji}$ indicating the number of times that player~$j$ has lost to player~$i$. This definition proves insightful for analysis as statistics governing the players may immediately be obtained. For example, the out-strength of node~$j$, $k_j^{\text{out}} = \sum_i w_{ji}$ describes the number of times that player~$j$ has lost and similarly their in-strength~$k_j^{\text{in}} = \sum_i w_{ij}$ gives their total number of wins, while the total number of matches in which they partook is simply the sum of these two metrics. The probability distributions of these quantities (alongside the corresponding complementary cumulative distribution function) are shown in the bottom panels of Fig.~\ref{fig:summaries} where  clear heavy-tailed distributions are observed, which is a common feature among networked descriptions of empirical social systems~\cite{Newman2010}. This suggests that the majority of snooker players take part in very few games and a select minority dominate the sport. 
	
	\subsection{Ranking Procedure}
	
	We now proceed to the main aim of this article, which is to provide a ranking scheme from which the relative skill of players may be identified. This is obtained through a complex networks approach which involves assigning each node~$i$ some level of importance~$P_i$ based upon their record across the entire network that is obtained via evaluating the PageRank score originally used in the ranking of webpage search results~\cite{Page1998} and more recently within the domain of sports~\cite{Radicchi2011, Mukherjee2012, Mukherjee2014, TennantBoxing17, Tennant20}. This procedure is mathematically described by
	\begin{equation}
		P_i = (1 - q) \sum_j P_j \frac{w_{ji}}{k_j^{\text{out}}} + \frac{q}{N} + \frac{1-q}{N} \sum_j P_j \delta\left(k_j^{\text{out}}\right),
		\label{eq:pageRank}
	\end{equation}
	which, importantly, depends on the level of importance associated with all other nodes in the network and as such is a coupled set of equations. Indeed, the first term within these equations describes the transfer of importance to player~$i$ from all other players~$j$ proportional to the number of games in which they defeated player~$j$, $w_{ji}$, relative to the total number of times the player lost, or their out-strength~$k_{j}^{\text{out}}$. The value of $q \in [0, 1]$ is a parameter (referred to as a damping factor in some literature e.g.,~\cite{Radicchi2009}), which controls the level of emphasis placed upon each term in the algorithm and has generally been set to 0.15 in the literature, which we follow here. The second term describes the uniform redistribution of importance to all players proportional to the damping factor, this allows the system to award some importance to nodes independent of their results. Finally, the last term contains a Kronecker delta, where $\delta(\cdot)$ is equal to one when its argument is zero and is otherwise zero, in order to give a correction in the case where nodes with no outdegree exist that would otherwise act as sinks in the diffusion process considered here. From the perspective of the sporting example studied here such nodes would represent undefeated players, none of which actually occur within the dataset. Therefore, while we show the final term in Eq.~(\ref{eq:pageRank}) for consistency with the existing literature, it can be neglected in the analysis to follow. 
	
	The main disadvantage of this approach is that, in general, an analytical solution to this problem proves elusive and as such we revert to numerical solutions, as in~\cite{Radicchi2011, Mukherjee2014}, by initially assigning each node an importance reciprocal to the network's size and iterating until convergence to a certain level of precision.  After the system of equations has reached its steady state in this diffusive-like process we proceed to rank the nodes by their corresponding importance scores.
	
	\section{Results}
	\subsection{All-time Rankings}
	Having implemented the system of equations given by Eq.~\eqref{eq:pageRank} using the network described in Sec.~\ref{subsec:network} we obtain a ranking of Snooker players over all time, the top 20 of which are shown in Table~\ref{tab:top20}. Immediately some interesting results appear. First, 18 of the players within this list are still competing in the sport which is indicative of two things--- first snooker players have considerably longer careers (regularly spanning over 30 years) in comparison to other sports and a second related point is that the current period of Snooker can be viewed as a \textit{golden-age} of sorts. It is important to note that these results may also be considered contrary to general opinion if one instead based their ranking upon the number of \textit{World Championships} (Snooker's premier tournament) a competitor has won where the top four ranked players are \textit{Stephen Hendry}~(7), \text{Steve Davis}~(6), \textit{Ronnie O'Sullivan}~(6), and \textit{Ray Reardon}~(6). However one must recognize an important factor which is vital to the algorithm used here, namely that in most of these cases the titles were won over a short period of time indicating that the prime years of these player's careers did not feature as much competition among other highly ranked players. For example, Hendry won his titles over a period of 10 years, Davis and Reardon in nine years each, whereas O'Sullivan has taken 20 years to amass his collection suggesting more competition between players he competed with and as such he is correspondingly given the highest rank of the four in our algorithm. 
	
	\begin{table*}[h!]
		\centering	
		\caption{\label{tab:top20}The top 20 players in Snooker's history.}
		\begin{tabular*}{\textwidth}{l @{\extracolsep{\fill}} lcclcc}
			\toprule
			Rank & Player & PageRank Score & In strength & Nationality & Start & End \\
			\midrule
			1 & John Higgins & 0.0204 & 899 & Scotland & 1992 & -\\
			2 & Ronnie O'Sullivan & 0.0201 & 843 & England & 1992 & -\\
			3 & Mark Williams & 0.0169 & 768 & Wales & 1992 & -\\
			4 & Stephen Hendry & 0.0164 & 818 & Scotland & 1985 & 2011\\
			5 & Mark Selby & 0.0149 & 643 & England & 1999 & -\\
			6 & Judd Trump & 0.0136 & 579 & England & 2005 & -\\
			7 & Neil Robertson & 0.0134 & 581 & Australia & 2000 & -\\
			8 & Steve Davis & 0.0129 & 761 & England & 1978 & 2014\\
			9 & Shaun Murphy & 0.0126 & 552 & England & 1998 & -\\
			10 & Jimmy White & 0.0116 & 650 & England & 1980 & -\\
			11 & Stephen Maguire & 0.0113 & 475 & Scotland & 1997 & -\\
			12 & Ali Carter & 0.0111 & 487 & England & 1996 & -\\
			13 & Peter Ebdon & 0.0110 & 520 & England & 1991 & -\\
			14 & Ken Doherty & 0.0110 & 523 & Ireland & 1990 & -\\
			15 & Barry Hawkins & 0.0105 & 475 & England & 2000 & -\\
			16 & Marco Fu & 0.0104 & 427 & Hong Kong & 1997 & -\\
			17 & Ding Junhui & 0.0103 & 436 & China & 2003 & -\\
			18 & Stuart Bingham & 0.0101 & 477 & England & 1996 & -\\
			19 & Mark Allen & 0.0100 & 444 & Northern Ireland & 2002 & -\\
			20 & Ryan Day & 0.0098 & 458 & Wales & 1998 & -\\
			\bottomrule
		\end{tabular*}
	\end{table*}
	
	Through this approach we identify \textit{John Higgins} to be the greatest Snooker player of all time, which is an understandable statement when one considers his career to date. Having already commented on the more competitive nature of the game since the late 1990s above, we note that Higgins has appeared in the top 10 ranked positions of the official rankings in 22 out of 25 years between 1995 and 2019 (20 of which he ranked in the top five positions) which is an impressive return in the circumstance. An interesting occurrence is also observed whereby the top three positions are filled by players who first competed professionally in 1992 (these three are in fact known as the \textit{Class of 92} within the Snooker community) and their frequent competition between one another over the pass 28 years is in itself helpful towards understanding their co-appearance as the highest ranked players.
	
	Two other interesting happenstances occur in the rankings shown here, namely the relatively low ranking of both \textit{Steve Davis} and \textit{Jimmy White} in spite of their large number of wins. This is again explained by them both experiencing their peak years in terms of winning tournaments in an era within which there were fewer successful players (it is worth noting that White is in fact infamous for having never won the World Championship despite reaching six finals) and as such their wins receive less importance in the algorithm.
	
\begin{figure*}
		\centering
		\includegraphics[width=0.85\textwidth]{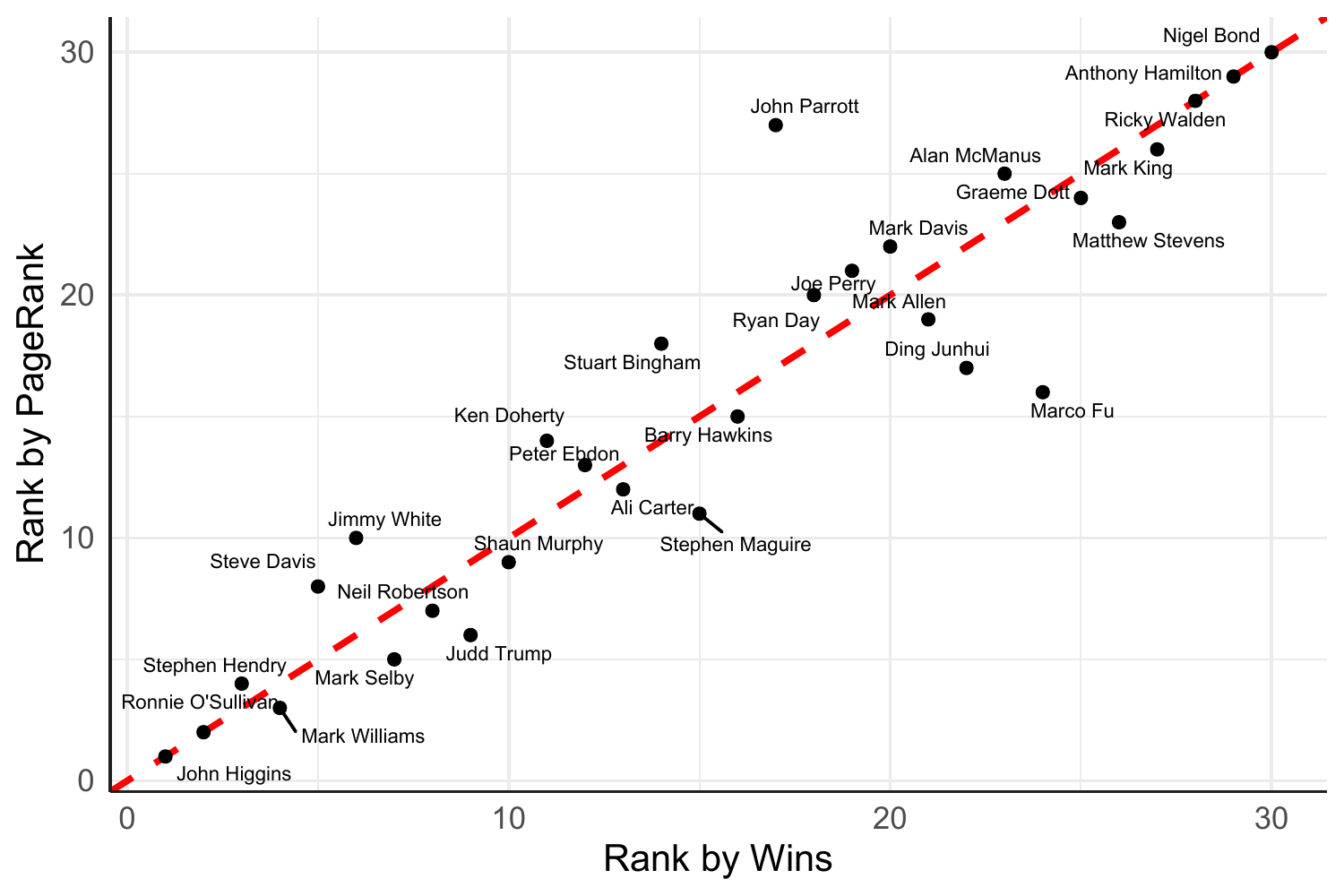}
		\caption{\textbf{Relationship between PageRank importance and number of wins.} The top 30 players over the full time period considered here (1968-2019) ranked by both the PageRank score and the number of wins obtained by the player, which interestingly offer an exact overlap of players. We see that the two measures are highly correlated with Spearman correlation $\rho = 0.932$, and Kendall $\tau = 0.793$.}
		\label{fig:all_time_rank}
	\end{figure*}

	One may observe from Table \ref{tab:top20} that there appears to be a strong correlation between a player's PageRank and their corresponding number of wins (their in-strength). This is further demonstrated in Fig.~\ref{fig:all_time_rank} where the relationship between the two is visualized. Quantitative comparison between the two measures provides a very strong correlation with a Spearman correlation $\rho = 0.932$ and a Kendall tau correlation of $\tau = 0.793$. While this correlation is strong it does indicate some disparity, particularly so in the case of the second measure, which demonstrates the subtle differences evident in the two approaches. Namely, the algorithm proposed in this article can identify the quality of opponent whom a player is defeating such that it captures more information than simply the result itself. This suggests that those players whom appear below the red-dashed line in Fig.~\ref{fig:all_time_rank} have had to obtain their career wins in more difficult contests and vice-versa for those above the line. This point further emphasizes the value in our proposed approach for identifying the top players. On the contrary, if instead the metric used was simply the number of wins the players would have incentive to enter as many tournaments possible, particularly those in which they had a better chance of less competitive games, in order to increase their number of wins.
	
	\subsection{Specific Seasons}
	While the analysis thus far has focused upon the entire breadth of the dataset we may also readily consider more specific time periods within which a ranking of players may be provided. Indeed this is particularly beneficial in the scenario where we consider sections of the data at a season level i.e., the annual representation of the game of Snooker. Taking this as our starting point we proceed to consider each of the 52 seasons in the dataset and calculate the importance of each player who features in said season using our proposed algorithm. Table~\ref{tab:number_ones} shows the highest ranked player in each of these seasons using three metrics --- PageRank, In-strength (number of wins), and the official rankings provided by World Snooker~\cite{worldsnooker}.
	
	\makeatletter
	\renewcommand\@makefnmark{\hbox{\@textsuperscript{\normalfont\color{black}\@thefnmark}}}
	\makeatother
	
	\begin{table*}[h!]
		\centering	
		\caption{\label{tab:number_ones}The highest ranked player each year from 1968 based upon PageRank, In-strength, and the official rankings from World Snooker. Note year denotes the calendar year in which a season begun, i.e., 1975 represents the 1975-76 season.}
		\begin{tabular*}{\textwidth}{l @{\extracolsep{\fill}} ccc}
			\toprule
			Year & PageRank & In-strength & World Snooker\\
			\midrule
			1968 & Ray Reardon & Ray Reardon & ---\\
			1969 & John Spencer & John Spencer & ---\\
			1970 & John Spencer & John Spencer & ---\\
			1971 & John Spencer & John Spencer & ---\\
			1972 & John Spencer & John Spencer & ---\\
			1973 & John Spencer & Graham Miles & ---\\
			1974 & John Spencer & John Spencer & ---\\
			1975 & Ray Reardon & Ray Reardon & Ray Reardon\\
			1976 & Doug Mountjoy & Doug Mountjoy & Ray Reardon\\
			1977 & Ray Reardon & Ray Reardon & Ray Reardon\\
			1978 & Ray Reardon & Ray Reardon & Ray Reardon\\
			1979 & Alex Higgins & Alex Higgins & Ray Reardon\\
			1980 & Cliff Thorburn & Cliff Thorburn & Cliff Thorburn\\
			1981 & Steve Davis & Steve Davis & Ray Reardon\\
			1982 & Steve Davis & Steve Davis & Steve Davis\\
			1983 & Steve Davis & Steve Davis & Steve Davis\\
			1984 & Steve Davis & Steve Davis & Steve Davis\\
			1985 & Steve Davis & Steve Davis & Steve Davis\\
			1986 & Steve Davis & Steve Davis & Steve Davis\\
			1987 & Steve Davis & Steve Davis & Steve Davis\\
			1988 & Steve Davis & Steve Davis & Steve Davis\\
			1989 & Stephen Hendry & Stephen Hendry & Stephen Hendry\\
			1990 & Stephen Hendry & Stephen Hendry & Stephen Hendry\\
			1991 & Stephen Hendry & Stephen Hendry & Stephen Hendry\\
			1992 & Steve Davis & Steve Davis & Stephen Hendry\\
			1993 & Stephen Hendry & Stephen Hendry & Stephen Hendry\\
			1994 & Stephen Hendry & Stephen Hendry & Stephen Hendry\\
			1995 & Stephen Hendry & Stephen Hendry & Stephen Hendry\\
			1996 & Stephen Hendry & Stephen Hendry & Stephen Hendry\\
			1997 & John Higgins & John Higgins & John Higgins\\
			1998 & Mark Williams & John Higgins & John Higgins\\
			1999 & Mark Williams & Mark Williams & Mark Williams\\
			2000 & Ronnie O'Sullivan & Ronnie O'Sullivan & Mark Williams\\
			2001 & Mark Williams & John Higgins & Ronnie O'Sullivan\\
			2002 & Mark Williams & Mark Williams & Mark Williams\\
			2003 & Stephen Hendry & Ronnie O'Sullivan & Ronnie O'Sullivan\\
			2004 & Ronnie O'Sullivan & Ronnie O'Sullivan & Ronnie O'Sullivan\\
			2005 & John Higgins & Ding Junhui & Stephen Hendry\\
			2006 & Ronnie O'Sullivan & Ronnie O'Sullivan & John Higgins\\
			2007 & Shaun Murphy & Shaun Murphy & Ronnie O'Sullivan\\
			2008 & John Higgins & John Higgins & Ronnie O'Sullivan\\
			2009 & Neil Robertson & Neil Robertson & John Higgins\\
			2010 & Shaun Murphy & Matthew Stevens & Mark Williams\\
			2011 & Mark Selby & Mark Selby & Mark Selby\\
			2012 & Stephen Maguire & Stephen Maguire & Mark Selby\\
			\arrayrulecolor{red}
			\midrule
			\arrayrulecolor{black}
			2013\footnote{Rankings changed to be based upon value of prize-winnings.} & Shaun Murphy & Neil Robertson & Mark Selby\\
			2014 & Stuart Bingham & Stuart Bingham & Mark Selby\\
			2015 & Mark Selby & Mark Selby & Mark Selby\\
			2016 & Judd Trump & Judd Trump & Mark Selby\\
			2017 & John Higgins & John Higgins & Mark Selby\\
			2018 & Judd Trump & Neil Robertson & Ronnie O'Sullivan\\
			2019 & Judd Trump & Judd Trump & Judd Trump\\
			\bottomrule
		\end{tabular*}
	\end{table*}
	
	\begin{figure*}
		\centering
		\includegraphics[width=0.85\textwidth]{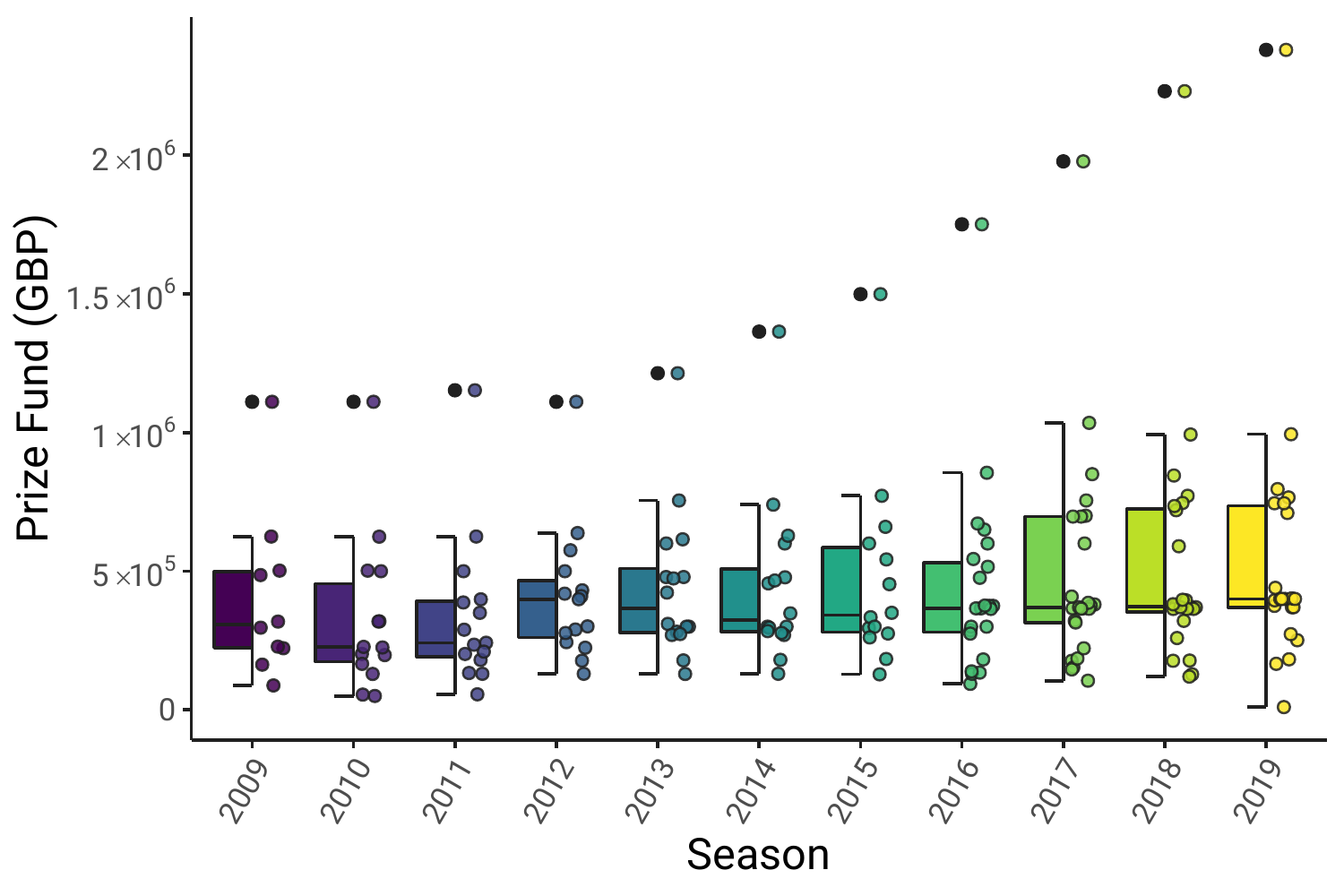}
		\caption{\textbf{Distribution of Prize funds within Snooker.} The total prize fund in each tournament for the last ten years within our dataset is shown. We highlight how the distributions have become more skewed in recent times, with the large outlier in each case representing the premier tournament in Snooker --- the World Championship. Note that the same outlier is represented twice each season i.e., by both the box and scatter plot.}
		\label{fig:prize_money}
	\end{figure*}
	
	This table offers an interesting comparison into how the best player in a given season is determined. For example, in the early editions we observe the first major benefit of our approach where due to there being no official rankings calculated to compare with, new inferences may be made in said years within which we identify \textit{Ray Reardon} and \textit{John Spencer} to be the dominant players. It is worth noting that in these early years there was a very small number of tournaments to determine the player ranking as evidenced in panels (a) and (b) of Fig.~\ref{fig:summaries}. Considering the years in which there are three measures to compare, we first comment on the general agreement between the PageRank and in-strength rankings for the first half of our dataset (in all years, aside from one, up to 1997 the two metrics agree on the number one ranking player) and in general offer a strong alignment with the official rankings. This is in agreement with our earlier statements regarding the level of competition present in these years such that the best player could be readily identified due to a lower level of competition. After this period however the level of agreement between the three metrics demonstrates considerably more fluctuation suggesting that the official rankings were not entirely capturing the true landscape of the game. This may have been one of the motivations for changing the official ranking procedure from the 2013-14 season to instead be based upon total monetary prize-winnings rather than the points-based system used previously. This has occurred, however, just as the monetary value of the largest competitions has increased significantly, as demonstrated in Fig.~\ref{fig:prize_money}, which can result in a skewed level of emphasis upon larger tournaments. Analysis of the seasons which have occurred since this change suggests that the problem has not been remedied by this alteration, demonstrated by the three measures all agreeing only twice in the seven subsequent seasons, whereas the two network based procedures on the other hand agree four times in the same period.
	
	\begin{figure*}
		\centering
		\includegraphics[width=\textwidth]{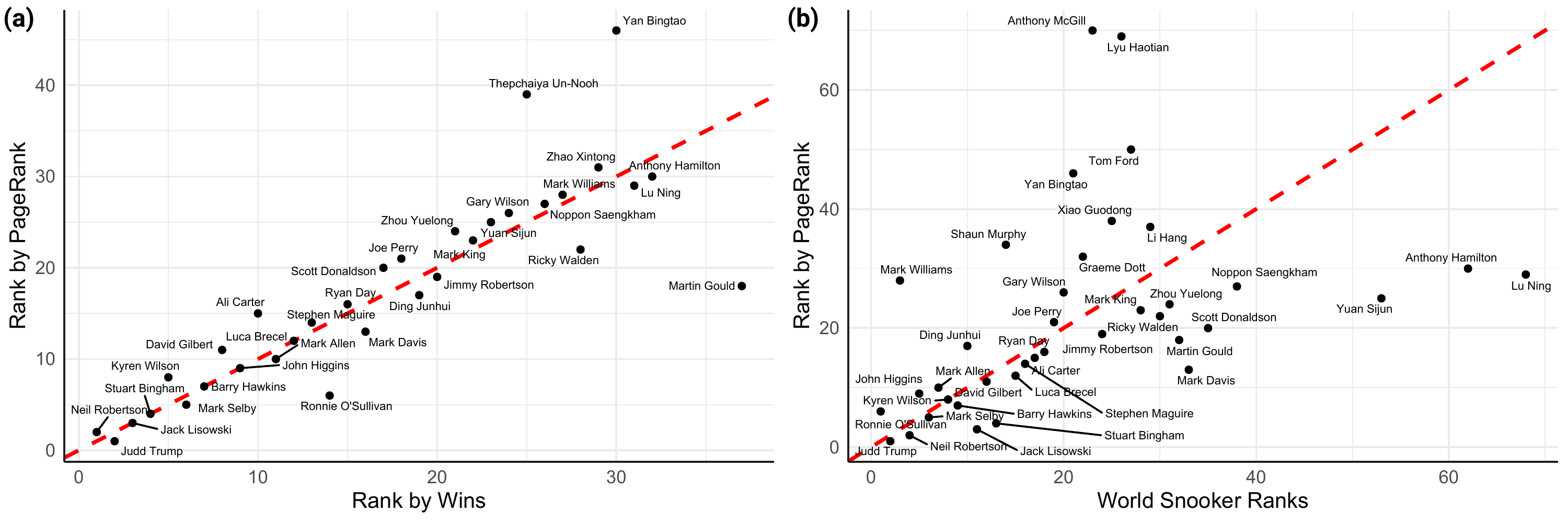}
		\caption{\textbf{Ranking algorithms in the 2018-19 season.} \textbf{(a)} The top 30 players from the 2018-19 season ranked by both the PageRank score and the number of wins obtained by the player, we see a strong correlation between the two metrics~(Spearman correlation $\rho = 0.917$, and Kendall $\tau = 0.792$). Note some players only appear in the top thirty ranks in one of the metrics. \textbf{(b)} Equivalent plot using the PageRank score and the official World Snooker rankings we now see a rather less correlated picture, particularly at larger ranks~($\rho = 0.635$, $\tau = 0.471$).}
		\label{fig:2018_ranks}
	\end{figure*}
	
	Figure \ref{fig:2018_ranks} shows a direct comparison between the top thirty ranked players obtained by the PageRank approach versus the two others. Specifically, we see in panel (a) that the PageRank and in-strength generally agree well with strong correlation between the two rankings ($\rho = 0.917$, $\tau = 0.792$). The equivalent features to those found in Fig.~\ref{fig:all_time_rank} are also relevant here, namely the location of a player relative to the red dashed-line is indicative of the types of matches they are taking part in. For example, we see that \textit{Ronnie O'Sullivan}, a player widely acknowledged to be among the greats of the game but has in recent years become selective in the tournaments in which he competes, being in self-proclaimed semi-retirement, is better ranked by the PageRank scores rather than his number of wins. This is in agreement with the idea that he generally focuses on the more prestigious tournaments (he took part in eleven in our dataset for this season, of which he won four) thus playing less matches but those he does play in tend to be against better players which is more readily captured by the PageRank algorithm. On the other end of the spectrum we have \textit{Yan Bingtao}, a young professional in the game, who's in-strength rank is stronger than his corresponding PageRank rank suggesting he has more wins against less prestigious opponents (he took part in eighteen such tournaments, unfortunately with no wins).
	
	The PageRank rankings are compared with the official World Snooker rankings in Fig.~\ref{fig:2018_ranks}(b) where we observe that the two sets of ranks are diverging considerably, particularly in the case of larger ranks. This has an effect on the corresponding correlation which is notably less than in the previously considered case ($\rho = 0.635$, $\tau = 0.471$). Both panels in Fig.~\ref{fig:2018_ranks} suggests that the PageRank is in some sense capturing the important features of the two alternative ranking metrics. In particular it appears that the algorithm is correctly incorporating the prestige of the tournament (shown by the good fit at higher ranks with the official rankings) while also more accurately capturing the performances of those players with lower ranks based upon their number of wins rather than their prize-winnings which can prove negligible in the case of not progressing far in tournaments.
	
	Lastly, the PageRank ranking scheme also, unlike traditional ranking systems, offers the advantage of being applicable across arbitrary time-spans. To demonstrate the potential of this we rank the top 10 players in each decade from the 1970s with the resulting players being shown in Table \ref{tab:top10_era}. These results highlight some interesting behavior in each decade, namely as commented on earlier, Steve Davis and Stephen Hendry are the highest ranked in the 1980s and 1990s respectively which are the periods in which they won all of their World Championships. The three highest ranked players shown in Table \ref{tab:top20}, the class of 92 - John Higgins, Ronnie O'Sullivan, and Mark Williams all feature in the top 10 rankings of three separate decades indicating their longevity in the sport and further justifying their positions in our all-time rankings.
	
	\begin{table*}[b]
		\caption{\label{tab:top10_era}The top 10 players by PageRank in different decades of Snooker's history.}
		\begin{tabular*}{\textwidth}{l @{\extracolsep{\fill}} ccccc}
			&&& Era &&\\
			\cmidrule{2-6}	
			Rank & 1970-1979 & 1980-1989 & 1990-1999 & 2000-2009 & 2010-2019\\
			\cmidrule{2-2}\cmidrule{3-3}\cmidrule{4-4}
			\cmidrule{5-5} \cmidrule{6-6}
			1 & Ray Reardon & Steve Davis & Stephen Hendry & Ronnie O'Sullivan & Judd Trump\\
			2 & John Spencer & Jimmy White & Ronnie O'Sullivan & John Higgins & Mark Selby\\
			3 & Alex Higgins & Terry Griffiths & John Higgins & Stephen Hendry & Neil Robertson\\
			4 & Doug Mountjoy & Dennis Taylor & Steve Davis & Mark Williams & John Higgins\\
			5 & Eddie Charlton & Stephen Hendry & Ken Doherty & Mark Selby & Shaun Murphy\\
			6 & Graham Miles & Cliff Thorburn & John Parrott & Ali Carter & Mark Williams\\
			7 & Cliff Thorburn & Willie Thorne & Jimmy White & Shaun Murphy & Barry Hawkins\\
			8 & John Pulman & John Parrott & Alan McManus & Neil Robertson & Ronnie O'Sullivan\\
			9 & Dennis Taylor & Tony Meo & Mark Williams & Ding Junhui & Stuart Bingham\\
			10 & Rex Williams & Alex Higgins & Peter Ebdon & Stephen Maguire & Mark Allen\\
			\bottomrule
		\end{tabular*}
	\end{table*}
	
	\subsection{Similarity of ranking metrics}
	
	With the aim of more rigorously quantifying the relative performance of the PageRank ranking scheme in comparison to both the official rankings and in-strength of players we consider the \textit{Jaccard similarity} of the approaches. This quantity is a metric used to determine the level of similarity between two sets $A$ and $B$ such that the Jaccard similarity of the two is given by
	\begin{equation}
		J(A,B) = \frac{|A \cap B|}{|A \cup B|},
		\label{eq:jaccard}
	\end{equation}
	where $|A \cap B|$ describes the number of players that appear in both set $A$ and $B$ while $|A \cup B|$ gives the total number of unique players appearing in both sets. The quantity itself is clearly defined in the range $[0,1]$ with zero indicating no similarity between the two sets and one suggesting two sets being equivalent.
	
	\begin{figure}
		\centering
		\includegraphics[width=\textwidth]{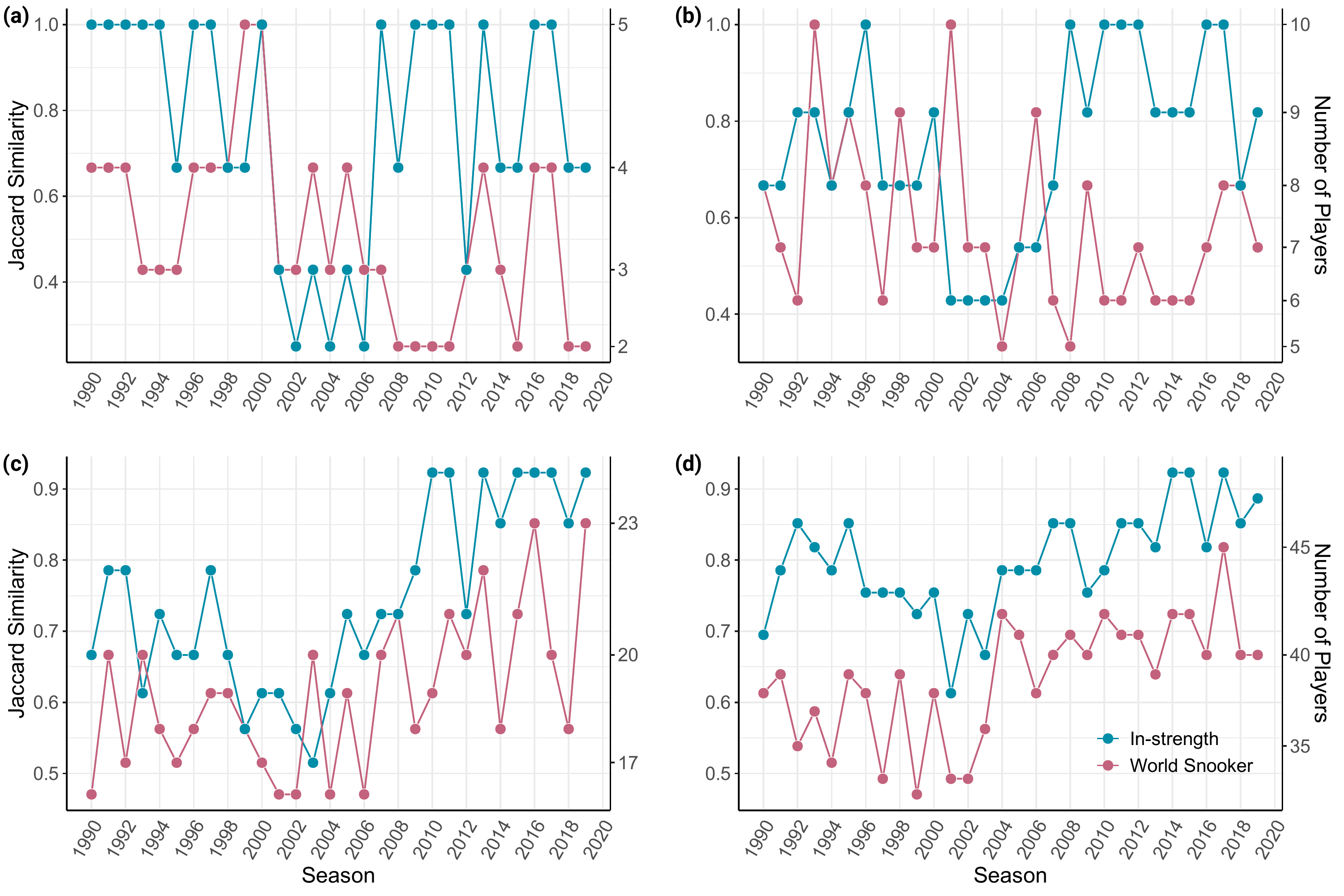}
		\caption{\textbf{Jaccard similarity between the ranking procedures.} \textbf{(a)} Similarity of the top five ranked players by PageRank to both the in-strength and official rankings, calculated via Eq.~\eqref{eq:jaccard}. The right-vertical axis describes the number of players overlapping in the two sets. Equivalent plots in the case of the top \textbf{(b)} 10, \textbf{(c)} 25, and \textbf{(d)} 50 ranked players are also shown.}
		\label{fig:jaccard}
	\end{figure}
	
	Figure~\ref{fig:jaccard} demonstrates this quantity in the case of the 5, 10, 25, and 50 ranks in each season. In each case the PageRank ranking is taken and compared with both the in-strength and official rankings. Significantly we again observe that the PageRank system offers strong, although not perfect, agreement to both alternative schemes in the case of the higher ranks (panels (a)-(c)). Furthermore the difference between its performance in the case of larger ranks is again evident as shown by the smaller similarity with the official rankings when the top 50 ranks are considered in panel (d). These results provide rigorous justification regarding the use of our PageRank scheme in ranking players as it accurately captures the better ranks in both alternative metrics while also more fairly representing those players with lower ranks through their total number of winning matches in comparison to the use of their prize-money winnings.
	
	\subsection{Rank-clocks}
	
	To provide an insight into the temporal fluctuations of a player's ranking throughout their career we make use of a tool known as a rank-clock~\cite{Batty2006}. These visualizations are obtained by transforming the temporal ranks to polar coordiantes with the rank being represented by the radial component and the corresponding year described through the angular part. As such the temporal variation of a player's rank over their career is demonstrated by the clockwise trend in the rank-clock. Such graphical techniques have previously been considered within the sport of boxing as a tool to consider future opponents for boxers \cite{TennantBoxing17}.
	
\begin{figure}
		\centering
		\includegraphics[width=0.95\textwidth]{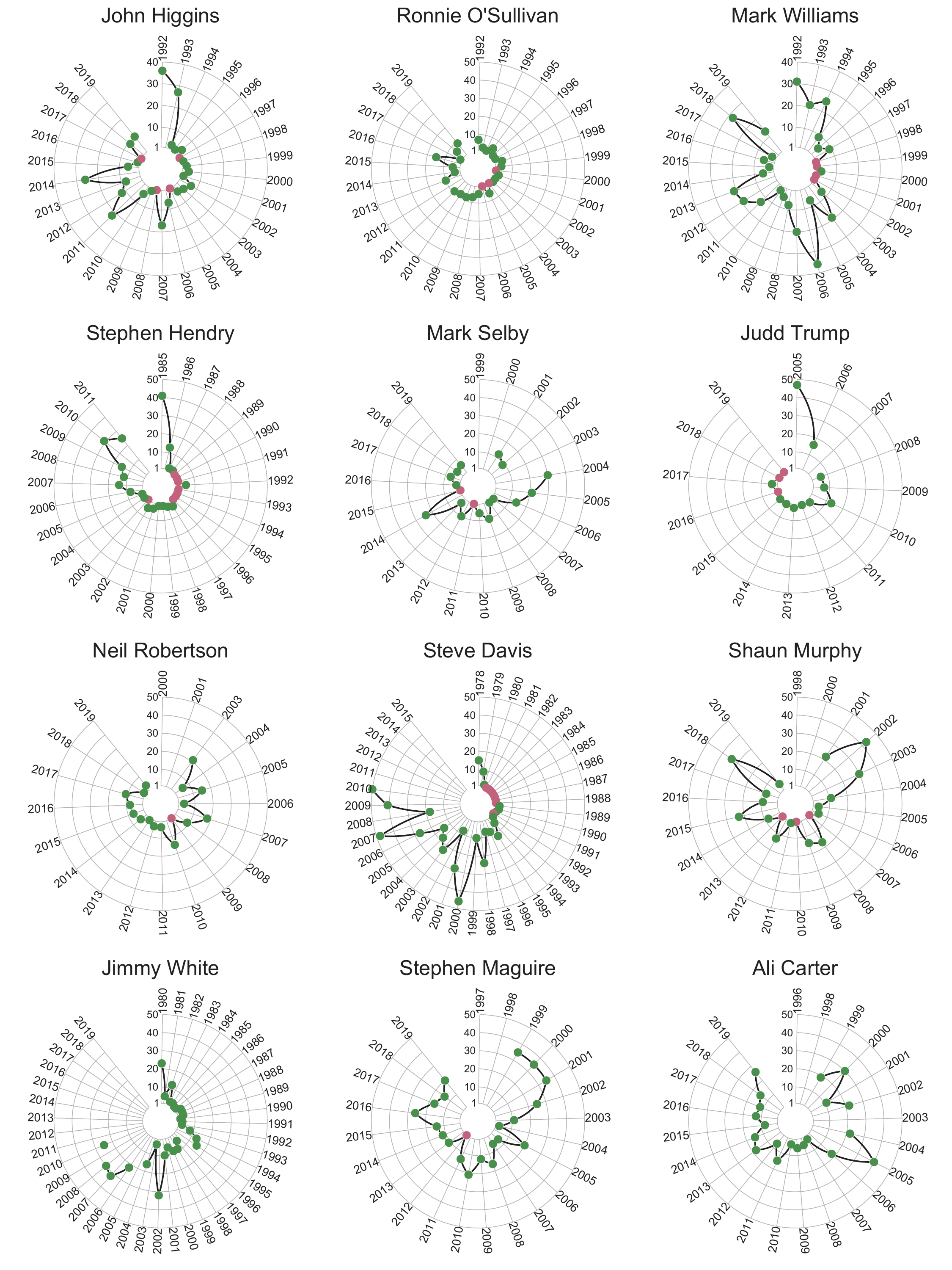}
		\caption{\textbf{PageRank ranks of the top 12 players in our dataset over the duration of their careers.} The temporal evolution of a player's PageRank ranking, calculated for each individual season, is shown through the radial element representing the rank while the polar coordinate represents the time evolution of the player's career. Discontinuities in the curves represents seasons in which the player finished outside the top 50 ranks. Seasons in which the player obtained a number one ranking are indicated by the purple dots.}
		\label{fig:rank_clock}
	\end{figure}

	Figure \ref{fig:rank_clock} shows the corresponding temporal fluctuations in the PageRank rankings of the top 12 players of all-time according to the preceding analysis. Each plot begins with the players first competing season before developing clockwise over the course of their career and finally ending with either the 2019-20 season rankings or their final competitive year. Importantly the outermost circle never goes beyond a rank of 50 and as seasons in which a player has a higher PageRank ranking than 50 are demonstrated by a discontinuity in the line. The top four ranked players again demonstrate their longevity in competitive performances with all four practically always being ranked within the top 40 performances each season (the exceptions are the 2012-13 season in which Ronnie O'Sullivan took an extended break and only competed in the World Championship --- which he won --- and the first season of Stephen Hendry's career).  
	
	These visualizations prove useful in quickly allowing one to obtain a perspective regarding the comparison of careers (and the direction in which they are going) for a selection of players. For example, a diverging curve in the later part of the clock indicates a drop in the competitive standards of a player which is clearly evident for the two who have already retired --- Stephen Hendry and Steve Davis --- and also Jimmy White who now frequently appears on the Senior tour within the game and only appears in tournaments on an invitational basis due to his pedigree in the game's history. The other extreme in which the players are performing stronger later in their career is evident by a converging curve in time, which is clearly the case for a number of the current crop of players including Judd Trump and Neil Robertson.
	
	\section{Discussion and Conclusions}	
	The human desire to obtain a definitive ranking of competitors within a sport is intrinsically flawed due to the large number of complexities inherent in the games themselves. Classical approaches to obtain rankings through methods such as simply counting the number of contests a competitor wins are unsatisfactory due to the lack of consideration regarding the quality of opponent faced in said contests. In this article, with the aim of incorporating these considerations into a ranking procedure, we provide further evidence of the advantages offered by network science in providing a more rigorous framework within which one may determine a ranking of competitors. Specifically, here we consider a network describing the matches played between professional snooker players during a period of over 50 years. We constructed this networked representation by observing the entire history of matches within the sport and demonstrated that this network has features consistent with previously studied social networks. 
	
	Having obtained this network structure, a much greater understanding of the competition that has taken place is possible by utilizing the network topology rather than simply considering individual matches. With this in mind we proceeded to make use of the PageRank algorithm, which has previously been shown to be effective in such scenarios for a number of sports~\cite{Radicchi2011, Mukherjee2012, Mukherjee2014, TennantBoxing17, Tennant20}, with the aim of obtaining a more efficient ranking system. The advantage of this approach is that it directly considers not only the number of contests a competitor wins but also directly incorporates the quality of opponent whom they are defeating and as such more accurately describe the performance of individual athletes. We show that this procedure is readily applicable to any temporal period one has data for thus allowing statements regarding the ranking at arbitrary points in time within the sport. Another important factor regarding our approach is that it requires no external consideration such as the points-based and monetary-based systems historically used in the official rankings. Furthermore we provide a quantification for the level of similarity between two different ranking schemes through the Jaccard similarity which offers validation of the benefits of our approach in capturing the ranks of players from various skill levels. A visualization tool in the form of the rank-clock is also introduced which offers a novel approach with which policy-makers within the sport of Snooker may quantify the success of  competitors over the temporal period of their careers. 
	\section*{Competing Interests}
	\noindent The authors have declared that no competing interests exist.
	\section*{Data Availability}
	\noindent All data and code used in this article is available at \url{https://github.com/obrienjoey/snooker_rankings}.
	
	\begin{acknowledgements}
		\noindent Helpful discussions with Edward Gunning are gratefully acknowledged. This work was supported by Science Foundation Ireland Grant No.~16/IA/4470, 16/RC/3918, 12/RC/2289P2, and 18/CRT/6049 (J.D.O’B. and J.P.G).
	\end{acknowledgements}
	
	\bibliography{snooker}
	
\end{document}